# On the Definition of Cyber-Physical Resilience in Power Systems

Reza Arghandeh[1*], Alexandra von Meier[1], Laura Mehrmanesh[1,] Lamine Mili[2],

[1]California Institute for Energy and Environment
Electrical Engineering and Computer Science Department
University of California-Berkeley
Berkeley, CA

[2]Eletrical and Computer Engineering Department
Virginia Polytechnic Institute and State University
Falls Church, VA
* Correspondent Author, Email: arghandehr@gmail.com

**Outline:**

1. Introduction
2. The "Pillars of Resilience" Concept
    2.1. From Risk Assessment to Resilience
    2.2. The Meaning of Resilience
    2.3. Going beyond Robustness
3. A Framework for Power System Cyber-Physical Resilience
4. Vulnerabilities in Power Systems
    4.1. Physical Vulnerabilities
    4.2. Cyber Vulnerabilities
    4.3. Cyber-Physical Vulnerabilities
5. Distributed Energy Resources in Microgrids, a Case Study for System Resilience
6. Conclusions and Future Work

**Abstract:**

Modern society relies heavily upon complex and widespread electric grids. In recent years, advanced sensors, intelligent automation, communication networks, and information technologies (IT) have been integrated into the electric grid to enhance its performance and efficiency. Integrating these new technologies has resulted in more interconnections and interdependencies between the physical and cyber components of the grid. Natural disasters and man-made perturbations have begun to threaten grid integrity more often. Urban infrastructure networks are highly reliant on the electric grid and consequently, the vulnerability of infrastructure networks to electric grid outages is becoming a major global concern. In order to minimize the economic, social, and political impacts of power system outages, the grid must be resilient. The concept of a power system's cyber-physical resilience centers around maintaining system states at a stable level in the presence of disturbances. Resilience is a multidimensional property of the electric grid; it requires managing disturbances originating from physical component failures, cyber component malfunctions, and human attacks. In the electric grid community, there is not a clear and universally accepted definition of cyber-physical resilience. This paper focuses on the definition of resilience for the electric grid and reviews key concepts related to system resilience. This paper aims to advance the field not only by adding cyber-physical resilience concepts to power systems vocabulary, but also by proposing a new way of thinking about grid operation with unexpected disturbances and hazards and leveraging distributed energy resources.

## 1. Introduction:

A widely dispersed asset, the electric grid has enormous impacts on people's lives. It is vital that the power grid can quickly recover with minimum damage after any intentional or unintentional outage. Severe weather is the leading cause of power outages in the United States, accounting for 87% of outages according to the 2013 report of the Executive Office of the U.S. President [1]. A recent congressional study estimates the cost of severe weather-related outages at an annual average of $25 to $70 billion [2]. It has been estimated that 90% of customer outages in the United States are related to distribution networks [3]. Moreover, distribution networks historically are behind transmission networks in terms of observability and monitoring system deployment. In the context of defining resilience, both transmission and distribution networks are taken into account. However, distribution networks need more attention in this area.

A "smart grid", one response to this situation, can strengthen the connection between information and communication technology (ICT) and advanced control systems. The synergy between physical power network components, communication network and cyber components may revolutionize grid efficiency and performance, but it also adds new cyber-access points. Increasing the number of access points increases the risk of physical damage by cyber-intruders. Moreover, interdependencies between electric transmission networks and distribution networks add more vulnerability to the power system as an interconnected entity. Power system vulnerability must therefore be evaluated from both the physical, the cyberspace and the interdependency perspective.

This paper aims to take a step forward not only by clarifying the concept of resilience, but also by proposing a new way of thinking about grid operation during unexpected disturbances, especially in distribution networks. Distribution networks experience fundamental behavioral changes with the growth of diverse distributed energy resources. This paper is a follow up to [4]. There is no clear and universally accepted definition of cyber-physical resilience for power systems. The first step in designing and operating resilient power systems is to clarify the definition of resilience. Current literature on power system resilience presents many conflicting and vague descriptions. Specifically,

1) The definitions of electric grid resilience in different publications do not always converge [5-7].
2) Service outages are well-studied in transmission networks, but not in distribution networks. Service outages in distribution networks have been increasing in recent years due to an aging grid and more frequent natural disasters. Moreover, the grid becomes more dynamic and complex in the presence of distributed energy resources [8].
3) The terms "robustness" and "resilience" are sometimes used interchangeably [9-11]. This is unfortunate, given that these are two different and sometimes mutually exclusive properties. Resilience hinges on flexibility and survivability in the face of unexpected events, while robustness implies resistance to change.
4) This paper aims to clarify differences between resilience and risk assessment objectives in the context of power systems.
5) The relationship between resilience and the concepts of reliability and stability in power systems requires more careful articulation [12, 13].
6) This paper highlights the role of distributed energy resources for enhancing grid resilience, especially in distribution networks.

As others have noted, cyber-physical network resilience (CPR) must be a temporal, agile, and holistic practice that makes the electric grid less vulnerable to outages and reduces the time of service recovery [14]. This paper defines resilience in power systems and provides a review of key related concepts, including robustness, hazards, vulnerability, risks, capacity and severity, focusing mostly on distribution networks. It aims to clarify the similarities and differences between these concepts – most notably robustness, a frequently misused word that has a specific and important meaning in the context of power system operating states. These definitional considerations provide the basis for this discussion of cyber and physical threats in power systems, and possible actions to mitigate these threats within a resilient infrastructure.

The paper is organized as follows: Section 2 focuses on the concept of resilience, and Section 3 presents a framework for understanding cyber-physical network resilience (CPR). Cyber-physical vulnerabilities in power systems are addressed in Section 4. In Section 5, distributed energy resources used for grid resilience enhancement are highlighted. Finally, Section 6 presents our conclusions and ideas for future work.

## 2. The "Pillars of Resilience" Concept

Before defining resilience in power systems, other concepts related to risk, hazard, vulnerability and robustness need to be clarified. They are the most commonly used terms in literature discussing system resilience. For risk concepts, the established analytic approaches for risk assessment can be of use in resilience analysis. For robustness concepts, in system engineering and control theory communities, "system robustness" is another concept used for system response in the presence of disturbances. Section 2.3 is devoted to comparing robustness and resilience concepts explicitly in power systems.

### 2.1. From Risk Assessment to Resilience

Risk and risk analysis are popular topics for scientists, engineers and politicians. While "risk" has different meanings in economics, business, politics and infrastructure, some common themes emerge. Risk assessment has been matured over the decades to analyze system damage probability following perturbations. This section aims to clarify the concept of risk in power systems to build a framework for a definition of resilience. Some literature uses risk assessment methodology for system resilience; that may not be a perfect approach. In infrastructure engineering, a discipline closely related to power systems, risk is assessed by two factors, the likelihood of an undesirable event and the consequence of that event [15].

**Definition 1**: *Risk is the possibility of an undesired event and its sequenced loss [15].*

In risk assessment, an event's occurrence likelihood and its consequences are characterized by probability distribution functions [16, 17]. For example, the risk of an overhead conductor line to ground fault is the probability of a line to ground short circuit and the fault consequences for customers. A common approach for risk quantification is the Triplet representation of the risk from Kaplan[17]. It focuses on the scenario, likelihood and consequence of the risk.

**Corollary 1**: The risk frequency and consequence are expressed in a set of probability distribution functions (PDF)[17]:

$$Risk_i = \left\{ \langle S_i, f_i(\varphi_i), g_i(\xi_i) \rangle \right\} , \ i = 1,2,... \tag{1}$$

where $S_i$, $p_i(\varphi_i)$, and $p_i(\xi_i)$ are risk scenario, likelihood PDF and consequence PDF for hazard *i*.

The Pressure and Release (PAR) risk model by Wisner [18] is another popular approach for risk modeling. The PAR model views disasters as the intersection of vulnerability and hazards. It has three determinates: hazard, capacity and vulnerability [19].

**Definition 2**: *A hazard is an event or set of events that is the source of potential damage. Hazards cause concerns for system owners and operators [20].*

In Corollary 1, the *i* subscript refers to possible hazards. The properties of hazard-generating sources are usually unknown, so they are represented with probability rules. The behavior of the system subjected to hazardous events can be probabilistic or deterministic. Hazards in power distribution networks can be natural phenomena such as vegetation, lightning, severe weather, and animals. Other types of hazards include malicious and terrorist attacks.

**Definition 3**: *Capacity is the ability of a system to adapt to imposed changes and moderate potential damage [21].*

For example in distribution networks, capacity is part of network planning for reserve capacity, conductor over-sizing and line redundancy. In generation, capacity can be in the form of spinning reserve for frequency droop control.

**Definition 4**: *Vulnerability is a condition or a process resulting from a given (natural or man-made) hazard and is defined as the joint conditional probability distribution of hazard likelihood, hazard potential impact and system capacity [22].*

**Definition 5**: *Severity is the statistical likelihood of hazards according to historical data.*

**Corollary 2**: Given the hazard *i* and $i \subseteq H$ where H is the set of possible hazards for the system, the vulnerability function will be:

$$Vul_i = f(\varphi|i) \times g(\xi|i) \quad (2)$$

where *f* is the conditional PDF for the hazard potential impact $\varphi$ relative to the hazard *i*, and *g* is the conditional PDF for the system capacity $\xi$ relative to the hazard *i*.

Drawing upon the definitions of the PAR model basics in Corollary 1, Corollary 3 describes the PAR mathematical formulation.

**Corollary 3**: The Pressure and Release (PAR) model of risk [23] is:

$$Risk_i = Vul_i \times Severity_i \quad (3)$$

where the $Vul_i$ is the vulnerability for hazard *i*.

Risk assessment based on hazard and vulnerability is a framework that presents risk in both a system behavior context and a physical characteristics context [24]. Wisner [18] provides a conceptual PAR risk framework (see Figure 1).

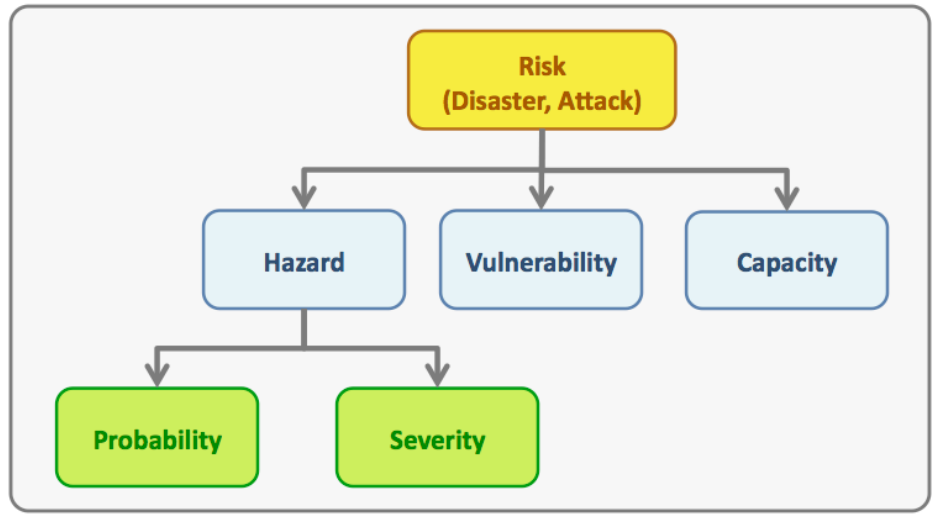

**Figure 1. Conceptual PAR risk framework [18].**

As Figure 1 shows, the damage of a system following the disturbances is a function of four different parameters, probability of the disturbance, severity of the disturbance, system vulnerability, and system capacity to absorb the disturbance.

Understanding the nature of a risk and its consequences and perturbations in a network is part of risk assessment procedure. While understanding risks is an important first step, the ultimate goal is to build power systems that are resistant to and nimble in the face of risks. Modern power systems need to adopt mechanisms to cope with risks and recover from outages quickly. The next section is focused on the concept of resilience in the context of the electric grid.

## 2.2. The Meaning of Resilience

The electric grid is a socio-ecological system with different spatial, temporal, and organizational parameters that are affected by policy, economy and society. Therefore, definitions of resilience in other disciplines can help us build an expressive definition of resilience in power systems. Definitions of resilience have evolved and expanded over the years. In 1973, Holling [25] defined resilience as the ability of a system to maintain its functionality and behavior after a disturbance. Gunderson et al. [26] modified the definition by adding buffer capacity for absorbing perturbations in a timely fashion. Walker et al. [27] extended the definition to include the ability to self-heal during disturbances. Kendra et al. [28] described "bouncing back from a disturbance" as a crucial aspect of resilience. The breadth of and number of definitions for "resilience" has increased significantly over the last decade, making it difficult to find a universal understanding of the term "resilience".

Table 1 presents different definitions for resilience in different disciplines. It shows how resilience definitions share similar concepts from different perspectives. The power systems community needs a tailored resilience definition that includes physical and cyber network characteristics and service outage consequences.

Table 1. Different Definitions of Resilience from Different Disciplines

| Discipline | Definition of "Resilience" | Ref |
|---|---|---|
| **Infrastructure Systems** | The ability to reduce the magnitude and duration of disturbances. It depends upon the system's ability to predict, absorb and adapt to disturbances and recover rapidly. | [29] |
| **Economic Systems** | The response to hazards that enables people and communities to avoid some economic losses at micro-macro market levels. It is the capacity for the enterprise to survive and adapt following market or environmental shocks. | [30] |
| **Social Systems** | The ability of a community to withstand stresses and disturbances caused by social, political and economic changes. | [31] |
| **Organizational Systems** | The ability of an organization to identify risks and to handle perturbations that affect its competencies, strategies and coordination. | [32] |

Resilience is especially critical immediately following an event that challenges system performance and functionality. Such events are given various names by different authors from various disciplines. A hazard, Definition 2 in this paper, is one such name for these events. Table 2 lists some of the other labels used. These terms describe consequences of rapid changes both in the environment and in system operation that are caused by system/component failures, attacks and natural disasters. From this point forward in the paper, "disturbing events" will include all of the terms in Table 2 and other similar terms.

Table 2. Different terms for "Disturbing Events" used in system resilience literature.

| Term | Ref | Term | Ref | Term | Ref |
|---|---|---|---|---|---|
| Perturbations | [33] | Losses | [34] | Anomalies | [35] |
| Disturbances | [35] | Adversity | [36] | Threats | [37] |
| Disruptions | [38] | Emergency | [39] | Shocks | [40] |
| Events | [41] | Changes | [42] | Hazards | [43] |

**Definition 6**: *The resilience of a system presented with an unexpected set of disturbances is the system's ability to reduce the magnitude and duration of the disruption. A resilient system downgrades its functionality and alters its structure in an agile way.*

Cyber-physical resilience assessment is often based on risk assessment [44, 45], which may not be the best approach for providing a system with a given degree of resilience. Risk assessment is the likelihood of failures in a probabilistic language. Resilience is about mitigation of unexpected failures, regardless of the failure's likelihood. Resilience assessment depends on the temporal dimension of potential disturbances and mitigating actions. Resilient structures find strategies to keep the backbone of the system intact. However, risk assessment centers around the probability of hitting a system's weak points.

In terms of a system's response to disasters, attacks and failures, risk assessment is a general framework to evaluate damage to the system performance and functionality. However, resilience is a quick reaction to damage and attacks with the goal of maintaining system functionality. The risk assessment goal is situational awareness and diagnostics. Resilience is taking one step forward while taking quick actions to maintain system functionality.

In resilience operations, response time and service availability are key. In the next sections, a more refined definition of resilience for power systems is presented.

## 2.3. Going Beyond Robustness

In the wake of unprecedented disasters and attacks, robustness and resilience have become buzzwords in many disciplines, including biology, ecology, sociology, systems engineering and infrastructure engineering. The traditional definition of resilience in systems engineering is the capacity for fast recovery after stress and for enduring greater stress [46]. In systems engineering, resilience includes maintaining system functionality following disturbances. Robustness, on the other hand, refers to the ability of a system to resist change without losing stability [4]. A more generic definition of robustness is:

**Definition 7**: *Robustness is the ability of a system to cope with a given set of disturbances and maintain its functionality.*

Robustness and resilience belong to two different design philosophies. Robustness is concerned with strength, whereas resilience is concerned with flexibility. When a robust grid is attacked, it may break like an oak tree in a storm. When a resilient grid is attacked, it can bend and survive like a reed in a storm [4]. From a systems engineering point of view, absolute robustness can actually lead to fragility.

In some disciplines like social systems and organizational systems, the term resilience is similar to the term robustness. In infrastructural systems, and especially in power systems, however, the terms robustness and resilience are more distinct; this is due to power systems' structure and function centering around conductor lines delivering electric power to a certain area within specific voltage and frequency ranges.

**Remark 1**: The more an infrastructure network is designed to be robust against one set of disturbances, the more fragile it is when faced with a different set of disturbances. Therein lies the fundamental connection and conflict. Extreme robustness actually leads to fragility. Power systems are usually robust enough to withstand one contingency (N-1) or two contingency (N-2) events, where N stands for the number of system buses. However, beyond that, they are generally vulnerable. Moreover, the term robustness is usually used with specific assumptions for protection system operation under pre-defined operational ranges for voltage and loading.

**Remark 2**: Robustness is usually embedded in the system's design, whereas resilience is typically integrated into the system's operational components like its control system. Robustness is defined against specific threats to the system. For example, distribution poles have to withstand earthquakes and wind speeds up to a certain level of structural stress and strain. System robustness requires stronger coupling between network components, like replacing overhead lines with underground cables. Resilience, on the other hand, demands flexibility, adaptability and agility. Dynamic system components like loads and distributed generation force sudden changes in system behavior. Resilient power systems know how to reroute electricity to customers using alternative paths and alternative local sources during natural disasters.

Robustness in the enterprise world is more focused on asset utilization, whereas resilience centers around service quality. Robustness is embedded in the system architecture design; resilience is more concerned with system operation. Robustness can be a passive approach for system security. Distribution pole hardening and putting cables underground are examples of passive system security enhancement. Resilience, on the other hand, is an active approach with real-time reactions to disturbances. Resilience can mean a set of real-time switching and islanding actions. Resilience can involve explicitly partitioning the grid into different sub-networks (microgrids). Robust electric grid networks try to maintain system functionality by damping perturbations. Resilient networks, on the other hand, rely on interdependencies to withstand perturbations. Therefore, multiple couplings between network components are crucial in resilient systems. Table 3 compares robustness and resilience against different criteria.

Table 3. Robustness vs. Resilience in Power Systems

| Criteria | Robustness | Resilience |
| --- | --- | --- |
| **Application** | Network hardening | Network flexibility |
| **Enterprise Focus** | Utility Assets | Utility Services |
| **Value Proposition** | Design | Operations |
| **Security Approach** | Passive | Active |
| **Network Preference** | Isolated | Interdependent |
| **Network Coupling** | Loose | Tight |

Reliability and stability are two more explicit power systems concepts that pre-date the terms "robustness" and "resilience" [4]. Reliability and stability are well-studied concepts in power systems. Similarities and dissimilarities between them and the terms robustness and resilience can inform future cyber-physical resilience studies. Reliability in power systems is the ability of grid components to meet all consumers' demand for electricity with acceptable power quality. The concept of reliability is also used in industrial and systems engineering and is accompanied by statistical and probabilistic approaches that characterize system performance after predicted and unpredicted failures.

**Definition 8**: *Reliability is the ability of the power system to deliver electricity to customers with acceptable quality and in the amount desired while maintaining grid functionality even when failures occur [47, 48].*

Discussions of resilience often center around a system survivability that leverages load shedding, generation outages, and other actions. Reliability is a measure of the system's ability to serve all loads. The system's ability to serve loads is traditionally referred to as service availability, which falls under the power systems definition of reliability. Reliability indices are usually expressed in terms of the probability of load loss [4]. The loss of load probability is expressed in days per year. The basic mathematical definition of reliability is presented in the next corollary. Reliability is primarily concerned with the risk of service interruption or device failure.

**Corollary 4**: The device failure at a random time *T>0* has the cumulative failure distribution function *F(t)*, probability density function *f(t)* and reliability *R(t)* as follows:

$$F(t) = P(T \leq t) \qquad (4)$$
$$R(t) = 1 - F(t) \qquad (5)$$

The next concept to clarify is the term "stability". Generally, stability is the system's ability to tolerate small perturbations. The small perturbations often come from uncertainties in measurements and system models. The concept of stability also comes up in control theory [49] and robust state estimation [50]. The general definition of stability is as follows:

**Definition 9**: *Stability is the ability of a system to remain intact after being subjected to small perturbations [51].*

In power systems, stability for a given initial operating condition means the system will regain operation equilibrium state after small perturbations. Stability is focused on the system equilibrium point. However, the concept of robustness in power systems goes beyond stability. In order to be robust, the electric grid has to be stable in the face of small perturbations as well as major equipment failures, man-made attacks, and natural disasters [51].

The previous sections built a foundation for defining resilience in power systems and terms that overlap with resilience, such as robustness, stability and reliability. The next section presents a definition of electric grid cyber-physical resilience.

## 3. A Framework for Power System Cyber-Physical Resilience

Understanding the nature of risk, its sources and its consequences is a major goal of risk assessment for a system. In power systems, in addition to the need for risk assessment, there is a need for actions performed in a timely manner to protect system functionality against risks, rapid changes and threats. Power systems are continually exposed to changing environmental and operational conditions caused by internal and external factors. The definition of resilience for power systems should be more holistic, rigorous and dynamic than what is encompassed by the term "risk assessment". Moreover, the electric grid is a complex, large scale and physically connected system with strong interdependencies between its components. A steady supply of electricity is vital for critical loads and facilities. However, continuous electricity delivery following natural and man-made disasters cannot be ensured without prioritizing loads and resources in response to disturbances. Power system resilience includes the survivability of the system and the system's ability to absorb the disturbances without losing its functionality. The following is a proposed definition for power system cyber-physical reliance.

**Definition 10**: *Power system cyber-physical resilience is the system's ability to maintain continuous electricity flow to customers given a certain load prioritization scheme. A resilient power system responds to cyber-physical disturbances in real-time or semi real-time, avoiding service interruptions. A resilient power system alters its structure, loads, and resources in an agile way.*

Power system cyber-physical resilience centers around the system's ability to recognize, adapt to, and absorb disturbances in a timely manner. Resilient system operation focuses on monitoring the system's boundary conditions to detect disturbances and adjusting control actions accordingly. Continuously monitoring the system creates a situational awareness for assessing risk, and supports system flexibility to mitigate disturbances. Power system resilience includes understanding the system's boundary conditions and their changes during disturbances [32].

Resilience is the system's ability to endure disturbing events in two ways: by absorbing disturbances ("absorbing potential") and by recovering from disturbances ("recovery potential"). Resilience implies that the system can absorb disturbances, adapt to the new parameters and recover fast enough to mitigate the effects of the disturbing event.

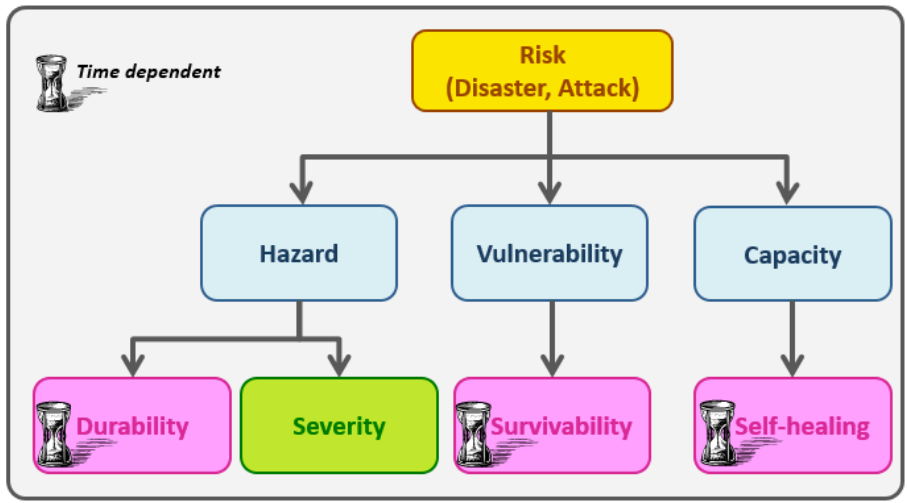

Figure 2. Comparing a resilience framework to a PAR risk analysis framework.

Comparing a resilience framework to a risk assessment model (Figure 1), we can see that power system resilience goes beyond the PAR model. First of all, in addition to knowing the severity of the hazard, one must know how long the system is being exposed to the hazard. Second, the probability of a disturbance is not a crucial factor in resilience, unlike in risk assessment. E.g., a longer duration storm causes more damage to the grid and requires more of a real-time resiliency response, whereas a merely more likely storm does not. This can be illustrated by considering how the impact varies according to the amount of time that a tree branch is touching an overhead line. The longer the branch lies on the conductor, the longer the short circuit current the system experiences.

As mentioned above, power system resilience is the electric grid's ability to survive disturbances. Resilience in power systems depends on the reaction time following a disturbance that maintains service availability. To modify the PAR model for grid resilience, the system vulnerability in the time domain is changed to system survivability. Similarly, system capacity for resilience is based on the self-healing characteristics of the network. Durability, survivability, and self-healing are time-dependent factors for power system resilience.

Resilience assessment requires knowledge of the power system's dynamic behavior and the system's flexibility in accommodating sudden changes without a tremendous decline in its performance. Therefore, a resilience assessment framework starts with system identification and model validation. Network topology, physical characteristics, operational constraints and dynamic behaviors are established in the system identification step. Topology detection and state estimation are integrated into the system identification process.

The next step is system vulnerability analysis. The next section reviews different vulnerabilities in power systems. Due to the randomness of disturbing events, their consequences are presented according to their likelihood in probability language. As the consequences of disturbing events are time-dependent, the temporal dynamics of disturbing events must be considered in resilience assessment [15]. In addition to the disturbing events' consequences, the system's adaptability and its recovery speed are crucial time-dependent factors that must be taken into account. Hence, vulnerability analysis includes the system response before, during and after disturbances. Vulnerability assessment is a continuous process; the evaluation of disturbing events and the consequences of the system's response to the events is ongoing.

The other component of the system resilience framework is resilient operation. The ultimate goal of a resilient system is to maintain system functionality after disturbing events. Resilient operation control defines new settings and equilibrium points for system operation. It has two main components, recovery potential and absorbing potential. These potentials are embedded in the resilience operation settings. Figure 3 depicts our proposed resilience framework for power transmission and distribution networks.

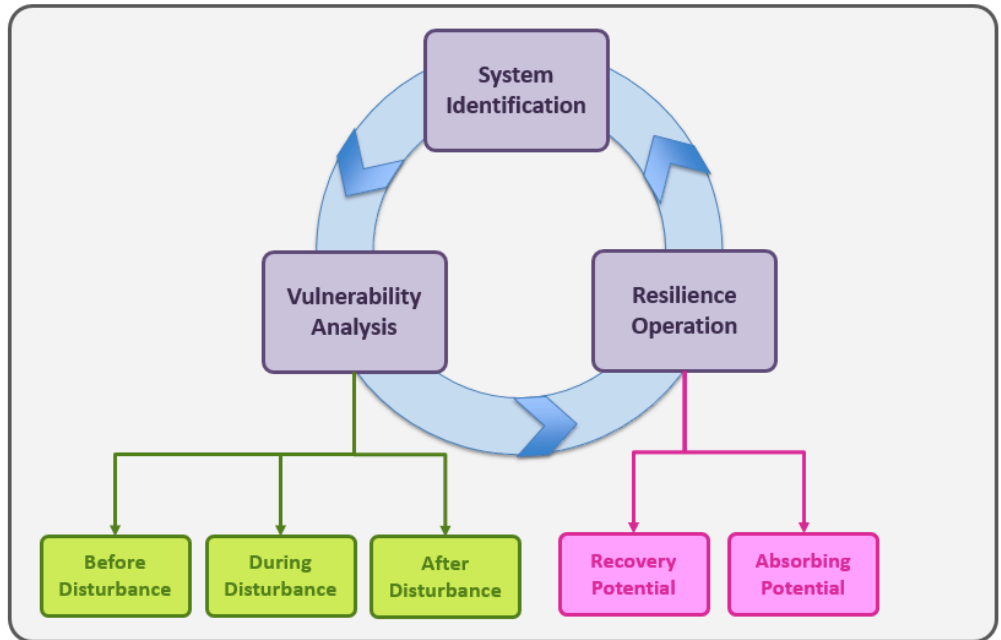

**Figure 3. Our Resilience Assessment Framework for power systems [52].**

Prior literature defines the absorbing potential as the degree to which a system can absorb the consequences of disturbing events [53]. The disturbance absorption in power systems depends on the components' design characteristics, the system topology, the control philosophy, and the protection coordination.

The recovery potential is the system's ability to alter itself in undesirable situations by recognizing disturbing events and reorganizing itself [19]. A quick return to normal operation or a restorative operation state is an important part of the recovery potential. The next section reviews common cyber-physical vulnerabilities, summarizing the current discourse in power system resilience operations.

### 4. Vulnerabilities in Power Systems

Power transmission and distribution networks are greatly dispersed and highly complex engineering systems with different degrees of connectivity. One of the key issues is that the dynamic electricity supply and demand balance needs to be maintained in real-time. Natural disasters, severe weather conditions and attacks make reliable operation a very difficult task. Electricity transmission and distribution networks as cyber-physical systems are a combination of physical grid components, sensors, communication devices, databases and software. Therefore, disturbing events in power systems can be organized into: 1) events in the physical grid components, grid structure and sensors, 2) events in the cyber infrastructure, software applications and data communication and 3) correlated events in power system components that have both cyber aspects and physical aspects, like control systems and state estimation systems.

### 4.1. Physical Vulnerabilities

Physical vulnerabilities are primarily due to the disruption of aerial distribution and transmission lines during and after severe weather. Faults caused by contact between conductors and ground are the source of circuit breakers locking out, safety hazards and fires. The second most vulnerable components are transformers. Hardening the distribution lines is one approach for preventing or mitigating the catastrophic effect of weather-related disruptions. Structurally reinforcing towers and poles is one effective way to increase robustness [54]. Vegetation management is crucial for preventing faults, especially in distribution networks. It is worth noting here that almost 90% of customer outages in the United States are related to distribution network problems [3].

In risk assessment studies, a common practice for determining infrastructure physical vulnerability is performing the fragility curve estimation [55]. This estimation method can also be used in resilience assessment. Han et al. [56] used data from a utility on the Gulf Coast to estimate fragility of overhead lines as a function of wind speed. Vickery et al. [57] introduced a curve-fitting technique for modeling structural damages. Francis et al. [58] presented underground lines' fragility curves.

There are extensive studies on the impacts of storms and natural disasters on the electric grid. [59] is a study on storms in Florida and their impacts on infrastructure. A more recent study on storm impacts on the grid and related state level legislation is presented in [60].

### 4.2. Cyber Vulnerabilities

Cyber-attacks and intrusions have been on the rise in recent years all over the world. As our power grid has gotten smarter, its components' communication abilities and information technology sophistication levels have increased. Unfortunately, that has resulted in an increase in the number of intrusion access points. Cyber intrusions can divulge critical data and measurements and cause a Denial of Service (DoS). Malicious commands and measurement injections can lead to widespread damage.

**Remark 4**: Cyber-attacks can be classified into four categories: 1) Reconnaissance, 2) Denial of Service (DoS), 3) Command Injection, and 4) Measurement Injection [61, 62].

The Department of Homeland Security [64] and the National Institute for Standards and Technology [63] have published assessments of cyber vulnerabilities in engineering systems. Table 4 shows some of the aforementioned vulnerabilities in smart grid related cyber systems. This list provides an overview of potential cyber-attacks on smart grids and the appropriate mitigation activities. Ten et al. and Hahn et al. [65, 66] have analyzed different types of cyber attacks on smart grid monitoring and protection systems.

Table 4. Some of the Cyber Vulnerabilities in Smart Grids [63, 64]

| Category | Common Vulnerability |
| --- | --- |
| **Software Domain Vulnerability** | - Improper Input Data Validation<br>- Poor Code Quality<br>- Permissions and Access Control<br>- Cryptographic Issues<br>- Improper Software Configuration<br>- Software Maintenance Issues |
| **Access Domain Vulnerability** | - Permissions, Access and Privileges Control<br>- Incorrect Authentication<br>- Improper Security Configuration<br>- Access Policy and Procedures Issues<br>- Credentials Management |
| **Network Domain Vulnerability** | - Improper Network Configuration<br>- Weak Firewalls<br>- Improper Network Component Configuration<br>- Network Audit and Monitoring Issues |

### 4.3. Cyber-Physical Vulnerabilities

The link between physical and cyber components in power systems makes it easy for cyber intrusions to cause physical damage to grid components [61]. Intelligent electronic devices (IEDs) and, in general, measurement devices with embedded communication and data processing, are used for different levels of control and protection in power systems. Control systems are where the cyber and physical systems come together. Cyber-physical vulnerability analysis should therefore start with the control systems, as suggested in industrial control security guidelines [64]. There are a number of research studies on cyber and physical interdependencies in control systems. Laprie et al. [14] analyze cascading failures that follow cyber-attacks on infrastructure control systems. Sridhar et al. [67] present a classification method for control system vulnerabilities for electric grid risk assessment. Qi et al. [68] propose a robust control algorithm for mitigating impacts of cyber attacks on power systems. Figure 4 illustrates the typical cyber-physical control system architecture for power systems.

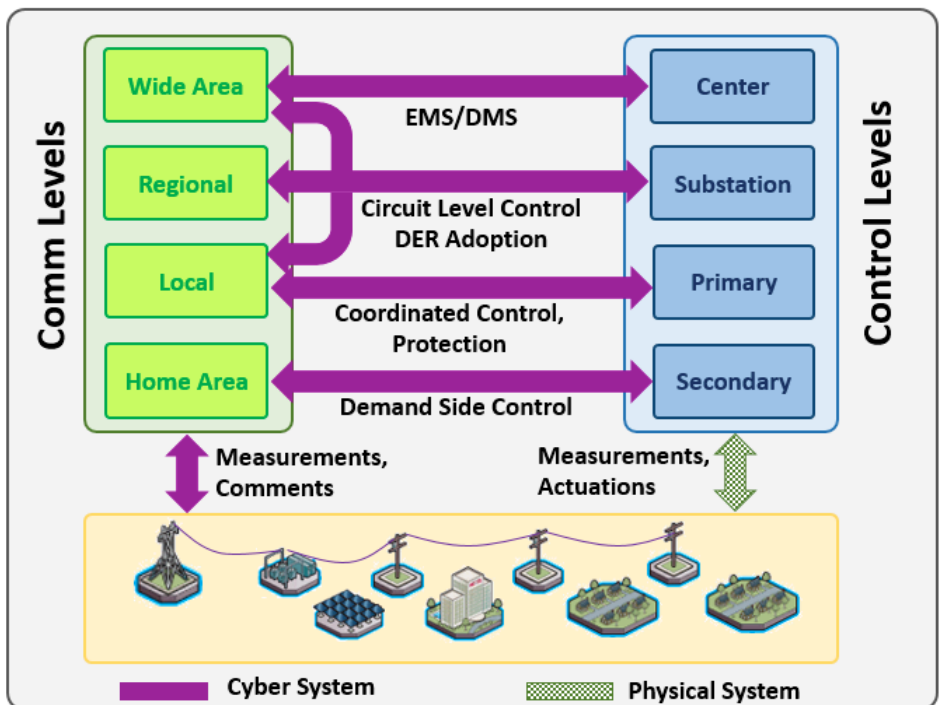

**Figure 4. A Typical Cyber-Physical Control System for an electricity grid.**

Figure 4 shows how measurement and control actuation signals are exchanged amongst physical network components. The solid arrows show the data path between different measurement and control components through communication lines. The communication links at the secondary control level of the distribution network and at the load level include advanced metering infrastructure (AMI) and home area network (HAN) technologies. The IEC 61850 standard is used for communication between coordinated control devices (voltage regulators, reclosers, breakers, etc.) and substations.

The Supervisory Control and Data Acquisition (SCADA) systems, AMI, and Distributed Energy Resource (DER) control systems play a major role in power system reliability services. These cyber-physical control and communication systems must be resilient against disturbing events and maintain grid performance under any circumstances. SCADA and AMI cyber-security issues have been explored by many researchers [69-72], but intelligent cyber-physical disturbance detection and the impact of DER on cyber-physical resilience has gotten less attention [73, 74].

## 5. Distributed Energy Resources in Microgrids, a Case Study for System Resilience

A resilient power system needs structural flexibility, modularity and distributed decision-making integrated with intelligent control and communication capabilities. Unfortunately, electric grid observability is a major challenge, especially in distribution networks. Power system resilience requires detailed knowledge of the system's behavior in three time scales: historical, real-time, and forecasting. Present monitoring systems do not typically have such extensive knowledge [75]. A well-designed monitoring system could be the backbone of system observability, capturing power system stresses, disturbances, and component failures that cause outages and service interruptions. In recent years, transmission networks have been equipped with time-synchronized phasor measurement units (PMUs) to monitor network stability. Distribution networks, however, are lagging behind transmission networks in this regard. Technologies like micro-synchrophasors and line sensors can help fill the gap [76].

To minimize the impacts of disturbances, resilient controllers can shed lower priority loads. Grid partitioning, suggested in earlier work, can be of use. Breaking distribution systems into islands [77], building AC and DC microgrids [8, 78], adopting more distributed energy resources (DER) [79], using intelligent power flow control systems [80] and creating distributed agent-based distribution network control systems [81] are all examples of efforts to find a general solution for contingency reduction, power flow control and flexible grid operation. The [8] suggests partitioning the grid into asynchronous sectors with DC interconnections. DC links let different partitions have quasi-independent frequency droop in response to disturbances. Grid partitioning lets system operators control power flows inside each segment (microgrid) and minimize stability issues and cascading failures in the larger network after a disturbance. In the extreme case, one segment collapses and other segments remain alive.

The advent of distributed energy resources (DER), such as renewable generation, electric vehicles, and controllable loads introduces great opportunities to help the grid survive and recover from extreme events. DER can provide local energy as well as more advanced ancillary services, even after extreme events. As illustrated in Figure 5, a solution can be achieved with a combination of intelligent contingency control on the transmission side along with DER adoption and microgrids on the distribution side. The interconnected transmission and distribution networks need sufficient numbers of measurement devices, coordinated control devices, and a communication network and hierarchical control system for data transmission and analysis. The resilient distribution management system maintains distribution networks' functionality with DER control, grid partitioning, load prioritizing, load shedding, and switching actions, along with new assumptions for equality and inequality constraints [52].

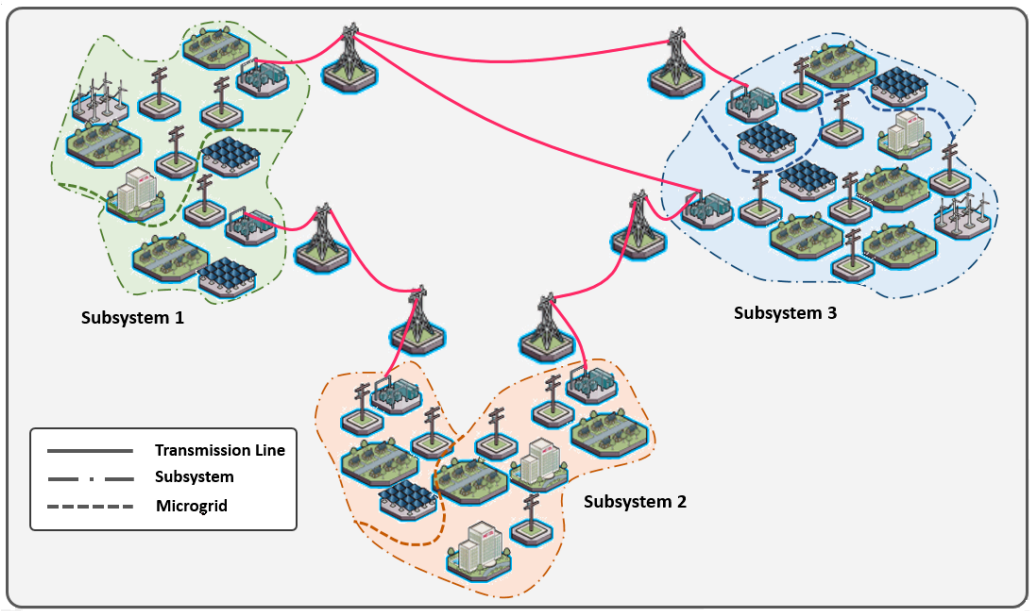

**Figure 5. Schematic of transmission and distribution networks with island operation capabilities.**

The intelligent distribution management system uses real-time control schema during emergency conditions. It takes advantage of a time-synchronized monitoring system for disturbance/failure detection via an updated alarm mechanism. With an intelligent distributed control system, DER, controllable loads, energy storage units, and switches can all participate in making the grid more resilient during and after disturbances. The current state of DER operation is the result of current standards and interconnection agreements that were developed when the penetration of distributed resources was low. Given the significant number of distributed resources that now exist at many utilities, and the forecasted growth of distributed resources, it is prudent to explore whether or not utilities could further leverage these resources in response to extreme events.

The IEEE1547 standard and Rule 21 from the California Energy Commission are initial efforts to regulate the interconnection, operation and measurement requirements for distributed energy resources. These regulations can be used as the basic standard for upcoming resilient grid operation frameworks with DER interconnections.

## 6. Conclusions and Future Work

Cyber-physical resilience for power transmission and distribution networks is an emerging discipline that requires further study. The comprehensive literature review in this paper shows that the field lacks a clear and custom-tailored definition for cyber-physical resilience in power systems, and in particular, in transmission and distribution networks. Previous work focuses mostly on risk assessment and robustness. The link between cyber and physical components in power systems has to be considered both when studying resilience itself and when creating a resilience assessment framework for power systems.

In this paper, a review of risk, hazard, vulnerability and severity in different disciplines is presented in order to provide a solid understanding of risk assessment and how it differs from resilience assessment. Since the term robustness is often misapplied, the authors then elucidate the differences between robustness and resilience in the context of power systems. Reliability and stability are two common concepts in power system operation and both concepts are related to robustness and resilience. This paper aims to illuminate the value of resilient distribution network operation and how it goes beyond reliability and stability.

A resilient system must go beyond risk assessment and carry out a set of actions in a timely manner to ensure adequate system functionality in the face of risks, sudden changes and threats. Power systems are continually facing variable operational conditions caused by internal and external factors. We posit that the concept of resilience in infrastructural networks must be centered on more holistic, rigorous and temporal analyses than those typically performed in traditional risk assessment.

This paper establishes a tailored and practical definition of cyber-physical resilience for power transmission and distribution networks. The next step is to use a probabilistic time domain framework to construct quantitative metrics for cyber-physical resilience operations in power systems. Disturbing events in power systems will be classified based on their resultant damage likelihood.